\documentclass{llncs}

\usepackage{a4,a4wide}
\usepackage{amsmath}
\usepackage{latexsym}
\usepackage{theorem}

\newtheorem{te}{Theorem}
\newtheorem{pr}[te]{Proposition}
\theoremstyle{definition}
\newtheorem{df}{Definition}
\newtheorem{ex}{Example}
\newtheorem{rk}{Remark}

\newcommand{\snot}{\mathit{not} \;}
\newcommand{\shead}{\mathit{head}}
\newcommand{\sbody}{\mathit{body}}

\title{Transfer of semantics from argumentation frameworks to logic programming. \newline A preliminary report}
\author{Monika Adamov\'a, J\'an \v{S}efr\'anek}
\institute{Comenius University, Bratislava, Slovakia,\newline  monika.adamova@gmail.com; sefranek@ii.fmph.uniba.sk}

\begin{document}

\maketitle

\begin{abstract}
There are various interesting semantics'  (extensions) designed for argumentation frameworks. They enable to assign a meaning, e.g., to odd-length cycles. Our main motivation is to transfer semantics' proposed by Baroni, Giacomin and Guida for argumentation frameworks with odd-length cycles to logic programs with odd-length cycles through default negation. The developed construction is even stronger. For a given logic program an argumentation framework is defined. The construction enables to transfer each semantics of the resulting argumentation framework to a semantics of the given logic program. Weak points of the construction are discussed and some future continuations of this approach are outlined.
\end{abstract}


\section{Introduction}

Relations between (extensions of) abstract argumentation frameworks and (semantics of) logic programs were studied since the fundamental paper by Dung \cite{dung} and since the times of other seminal paper \cite{bondarenko}. We can mention also, e.g., \cite{vermeir,prak,caminada,woltran,eggly,cf2woltran,osorio1,osorio2,osorio3,osorio4,osorio5,osorio6}.

Among typical research problems are, e..g., 
\begin{itemize}
\item a characterization of extensions of abstract argumentation framework in terms of answer sets or other semantics' of logic programs, 
\item a construction of new semantics of logic programs, based or inspired by extensions of argumentation frameworks,
\item encoding extensions in answer set programming.
\end{itemize}

Our main motivation is to transfer semantics' proposed in \cite{baroni} for argumentation frameworks with odd-length cycles to logic programs with odd-length cycles through default negation. According to our knowledge, only CF2 extensions of \cite{baroni}, were studied from different logic programming points of view, see, e.g., \cite{cf2woltran,osorio6}. In \cite{cf2woltran} an ASP-encoding of (modified) CF2 is presented and in \cite{osorio6} a characterization of CF2 in terms of answer set models is proposed.

Our goal is to propose some new semantics' of logic programs (we are primarily interested in a semantic handling of odd cycles through default negation) via transferring semantics' of argumentation frameworks (AD1, AD2, CF1, CF2). We propose a uniform method, which for a given logic program transfers arbitrary argumentation semantics to a semantics of the logic program. The method enables to define for a given logic program a corresponding argumentation framework. As next step, each semantics of the resulting argumentation framework is transferred to a semantics of the given logic program. 

This paper is structured as follows. Basics of SCC-recursive semantics of \cite{baroni} is sketched after technical preliminaries. Then, in Section 4, the core of the paper, a transfer of argumentation framework semantics' to logic program is described. A special attention is devoted to the problem of odd cycles in the Section 5. A representation of an argumentation framework $A$ by a logic program $P$ is described in Section 6. It is shown that for an arbitrary argumentation semantics holds that extensions of the original argumentation framework $A$ coincide with extensions of the argumentation framework constructed for $P$ using the method of Section 4.  Weak points of the construction are discussed in the paper. Some future continuations of this research are outlined in Section 7. Finally, related work is overviewed and main contributions, open problems and future goals are summarized in Conclusions.

\section{Preliminaries} \label{prelim}

Some basic notions of argumentation frameworks and logic programs are introduced in this section.

\paragraph{Argumentation frameworks}

An {\em argumentation framework} \cite{dung} is a pair $AF = (AR, atatcks)$, where $AR$ is a set (of arguments) and $attacks \subseteq AR \times AR$ is a binary relation. Let be $a, b \in AR$; if $(a, b) \in atatcks$, it is said that $a$ attacks $b$. We assume below an argumentation framework $AF = (AR, attacks)$.

Let be $S \subseteq AR$. It is said that $S$ is {\em conflict-free} if for no $a, b \in S$ holds $(a, b) \in attacks$. 

A set of arguments $S \subseteq AR$ attacks $a \in AR$ iff there is $b \in S$ s.t. $(b, a) \in attacks$. 

A conflict-free set of arguments $S$ is {\em admissible} in $AF$ iff for each $a \in S$ holds: if there is $b \in AR$ s.t. $(b, a ) \in attacks$, then $S$ attacks $b$, i.e. an admissible set of arguments counterattacks each attack on its members.

Dung defined some semantic characterizations (extensions) of argumentation frameworks as sets of conflict-free and admissible arguments, which satisfy also some other conditions.

A {\em preferred extension} of $AF$ is a maximal admissible set in $AF$. A conflict-free   $S \subseteq AR$ is a {\em stable extension} of $AF$ iff $S$ attacks each $a \in AR \setminus S$.

The {\em characteristic function} $F_{AF}$ of an argumentation framework $AF$ assigns sets of arguments to sets of arguments, where $F_{AF}(S) = \{a \in AR \mid \forall b \in AR \; (b \; \text{attacks} \; a \Rightarrow S \; \text{attacks} \; b \}$.

The {\em grounded} extension of an argumentation framework $AF$ is the least fixed point of $F_{AF}$ ($F_{AF}$ is monotonic).

A {\em complete extension} is an admissible set S of arguments s.t. each argument,  which is acceptable  with respect to $S$, belongs to $S$.

We will use a precise notion of a semantics of an argumentation framework.
A {\em semantics} of $AF$ is a mapping $\sigma_*$, which assigns a set of extensions to $AF$. Different indices in the place of * specify different semantics', e.g. preferred semantics, stable semantics etc. A set of extensions assigned by a semantics ${\mathcal S}$ to an argumentation framework $AF$ is denoted by ${\mathcal E}_S(AF)$.

\paragraph{Logic programs}

Only propositional normal logic programs are considered in this paper. Let $\mathcal L$ be a set of atoms. The set of default literals is $\snot \mathcal L = \{ \snot A \mid A \in \mathcal L \}$. A literal is an atom or a default literal. A rule (let us denote it by $r$) is an expression of the form
\begin{eqnarray}
A \leftarrow A_1, \dots, A_k, \snot B_1, \dots, \snot B_m; \; \text{where} \; k \geq 0, m \geq 0
\end{eqnarray}
$A$ is called the head of the rule and denoted by $\shead(r)$. \\
The set of literals $\{ A_1, \dots, A_k, \snot B_1, \dots, \snot B_m \}$ is called the body of $r$ and denoted by $\sbody(r)$. $\{ A_1, \dots, A_k \}$, called the positive part of the body, is denoted by $\sbody^+(r)$ and $\{ B_1, \dots, B_m \}$ is denoted by $\sbody^-(r)$. Notice that $\sbody^-(r)$ differs from the negative part $\{ \snot B_1, \dots, \snot B_m \}$ of the body.

A (normal) program is a finite set of rules. We will often use only the term program.

We will specify a transfer of an argumentation semantics to a logic program semantics in terms of sets of atoms derivable in the corresponding logic program. We follow the approach of Dimopoulos and Torres \cite{dt} in order to specify a notion of derivation in a normal logic program. The derivation should be dependent on a set of default literals. In the next paragraphs we will adapt some basic definitions from \cite{dt}.

An {\em assumption} is a default literal. A set of assumptions $\Delta$ is called a {\em hypothesis}. 
$\Delta^{\leadsto^{P}}$ is a set of atoms, dependent on (derivable from) $\Delta$ w.r.t. a program (set of rules) $P$; here is a precise definition:
 
Let $\Delta$, a hypothesis be given. 
$P_\Delta$ is the set of all rules from $P$, where elements from $\Delta$ are deleted from the bodies of the rules and $P^{+}_\Delta$ is obtained from $P_\Delta$ by deleting all rules $r$ with bodies containing assumptions. 
Then
$\Delta^{\leadsto^{P}} = \{ A \in \mathcal{L} \mid P^{+}_\Delta \models A \}$).

It is said that an atom $A$ is {\em derived} from $\Delta$ using rules of $P$ iff  $A \in \Delta^{\leadsto^{P}}$.

Stable model semantics of logic programs play a background role in our paper, so, we introduce a definition of stable model.
An interpretation $S = \Delta \cup \Delta^{\leadsto^{P}}$ is a stable model of $P$ iff $S$ is total interpretation \cite{dt}, where an interpretation is understood as a consistent set of literals.

\section{SCC-recursive semantics} 
\label{kap_cycles}

An analysis of asymmetries in handling of even and odd cycles in argumentation semantics' is presented in \cite{baroni}. We present only a sketchy view of their approach, for details see \cite{baroni}.

An argumentation framework may be conceived as an oriented graph with arguments as vertices and the attack relation as the set of edges.

\begin{ex}
Consider $AF = ( \{a, b, c \}, \{ (a, b), (b, c), (c, a) \})$. The graph representation of AF contains an odd-length cycle.

This example is often presented as a case o three witnesses and the attack relation is interpreted as follows: $a$ questions reliability of $b$, $b$ questions reliability of $c$, $c$ questions reliability of $a$.

Stable semantics does not assign an extension to such argumentation framework. However, there are two stable extensions for the case of four witnesses.

This asymmetry in semantic treatment of odd and even cycles motivated the research and solutions of \cite{baroni}. The same problem is present in a form also in other ``classical'' argumentation semantics proposed in \cite{dung}.
$\Box$
\end{ex}

A general recursive schema for argumentation semantics is proposed in \cite{baroni}.
Recursive semantics' are defined in a constructive way -- an incremental process of adding arguments into an extension is specified.

A symmetric handling of odd and even cycles is based on distinguishing components of graphs.

\begin{df}
Let an argumentation framework $AF=\langle AR,attacks \rangle$ be given. A binary relation of path equivalence, denoted by $PE_{AF} \subseteq(AR \times AR)$, is defined as follows.
\begin{itemize}
\item $\forall a \in AR, (a,a) \in PE_{AF}$,
\item $\forall a \not = b \in AR, (a,b) \in PE_{AF}$ iff there is a path from $a$ to $b$ and a path from $b$ to $a$.
\end{itemize}

The {\em strongly connected components} of $AF$ are the equivalence classes of arguments (vertices) under the relation of path-equivalence. The set of the strongly connected components of $AF$ is denoted by $SCCS_{AF}$.
\end{df}

We now can consider the set of strongly connected components as the set of vertices of a new graph. Consider components $C_1$ and $C_2$. Let an argument $a$ be a member of $C_1$ and $b$ be a member of $C_2$. If $a$ attacks $b$ (in AF), then $(C_1, C_2)$ is an edge of the graph of  strongly connected components (SCC-graphs). It is clear that this graph is an acyclic one.

Notions of parents and ancestors for SCC-graphs are defined in an obvious way.
Initial components (components without parents) provide a basis for a construction of an extension. We start at the initial component and proceed via oriented edges to next components. If we construct an extension $E$ and a component $C$ is currently processed, the process consists in a choice of a subset of $C$, i.e. a choice of $E \cap C$ (according to the given semantics -- the semantics specifies how choices depend on choices made in ancestors of $C$). A base function is assumed, which is applied to argumentation frameworks with exactly one component and it characterizes a particular argumentation semantics.

A notion of SCC-recursive argumentation semantics formalizes the intuitions presented above. SCC-recursive characterization of traditional semantics' is provided. Finally, some new semantics', AD1, AD2, CF1 and CF2, are defined in \cite{baroni}.

AD1 and AD2 extensions preserve the property of admissibility. However, the requirement of maximality is relaxed, so this solution is different as compared to the preferred semantics. 
An alternative is not to require admissibility of sets of arguments and and insist only on conflict-freeness. Maximal conflict-free sets of arguments are selected as extensions in semantics CF1 and CF2.
For details and differences see \cite{baroni}. ASP-encodings of AD1, AD2, CF1 and CF2 are presented in \cite{monika}.

\section{Transfer of argumentation framework semantics' to logic program}
\label{transAF}

We will build an argumentation framework over the rules of a logic program. Rules will play the role of arguments. An attack relation over such arguments will be introduced. After that some arguments (rules) are accepted/rejected on the basis of a given argumentation semantics. A corresponding semantics for logic program is introduced as a set of literals derivable from accepted rules (considered as arguments). Note that this method enables a transfer of an arbitrary argumentation semantics to the given logic program.

\begin{df}
\label{zaklDef}
Let a program $P$ be given. Then an {\em argumentation framework over} $P$ is $AF_P=\langle AR, attacks \rangle$, where

$AR = \{ r \in P \}$ and $attacks = \{ (r_1, r_2) \mid  A = head(r_1), body^+(r_1) = \emptyset, A \in body^-(r_2) \}$. 
$\Box$
\end{df}

\begin{ex}
Let be $P = \{ r_1:a \leftarrow; \; r_2:b\leftarrow \snot a.\}$. Then $\;attacks=\{(r_1,r_2)\}$ in $AF_P$.

If $P = \{ r_1:a \leftarrow \snot b.\; r_2:b\leftarrow \snot a.\}$, then  $attacks=\{(r_1,r_2),(r_2,r_1)\}.$
$\Box$
\end{ex}

Let us discuss the condition that the attacking rules do not contain positive literals in its body. A derivation of the head of a rule $r$ with non-empty $\sbody^+(r)$ from a hypothesis $\Delta$ is conditional: it depends on a derivation of positive literals in $\sbody^+(r)$. We constrain the attacking argument in the attack relation to the rules with non-empty $\sbody^+(r)$ -- it is recognizable on syntactic level  and it is appropriate for the representation of argumentation frameworks in logic programs presented in Section \ref{repre}.

But this design decision leads to some counterintuitive consequences in a general case. We will return to the problem below, after formal definitions.

We have defined an argumentation framework over the rules of a program $P$. Let's proceed towards derivations in $P$, based on an argumentation semantics.

Let a program $P$ be given, $AF_P$ be an argumentation framework over $P$. Consider a set of rules $R \subseteq P$, where $R$ is a conflict-free set of arguments of $AF_P$. It is obvious that $R$ could serve as  a basis of a reasonable derivation in the corresponding logic program. Only literals which do not occur as negated in the bodies of rules are in the heads of rules.

Notice that extensions of an argumentation framework over a program $P$ are sets of rules. That is expressed by a notion of rules enabled in a program $P$ by an argumentation semantics according to the following definition.

\begin{df}
\label{enabled}
A set of rules $R \subseteq P$ is {\em enabled} in a program $P$ by an argumentation semantics $S$ iff $R\in \mathcal{E}_S(AF_P)$. 
If $R$ satisfies this condition, it is denoted by $Rule\_in^P_S$ (or by a shorthand $Rule\_in$, if a given semantics and a given program are clear from the context).
$\Box$
\end{df}

A set of rules $R$ ($Rule\_in^P_S$) is enabled by $S$ according to Definition \ref{enabled}, if $R$ is an $S$-extension of $AF_P$. The following definition of a set of atoms consistent with a set of rules is important. It partially prevents some negative consequences of the decision  that attacking rules have empty positive part of the body. Inconsistent sets of rules cannot be derived because of checking consistency, see Definition \ref{derived}.  

\begin{df}
\label{consistent}
Let $M$ be an arbitrary set of atoms and $R \subseteq P$ be an arbitrary subset of a program$P$. 

It is said that $M$ is {\em consistent} with $R$ iff $\forall A\in M \; \neg \exists r\in R \; A \in body^-(r)$.
$\Box$
\end{df}

Now, a fundamental task is to point out a way from $Rule\_in^P_S$, rules enabled by an argumentation semantics to a corresponding set of atoms, i.e., to a semantics of the given logic program $P$. The set is denoted by $In\_AS_{\mathcal{S}}$, see the following definition.

\begin{df}
\label{in-as}
Let $AF_P$ be an argumentation framework over a program $P$, $\mathcal{S}$ be an argumentation semantics of $AF_P$ and $Rule\_in_{\mathcal{S}}$ is a set of rules   of $P$ enabled by the semantics $\mathcal{S}$. 

Then $In\_AS_{\mathcal{S}}$ is the least set of atoms $A$ satisfying the following condition:

$\exists r \in Rule\_in_{S}, head(r) = A, \forall b \in body^+(r): b \in In\_AS_{\mathcal{S}}$. 
$\Box$
\end{df}

Definition \ref{in-as} specifies how to compute $In\_AS$. First, for each $r \in Rule\_in_{S}$ s.t. $\sbody^+(r) = \emptyset$ and $\shead(r) = A$, $A$ is included into  $In\_AS$. After that is $In\_AS$ iteratively recomputed for all $r \in Rule\_in_{S}$ with non-empty $body^+(r)$. Notice that this is a process of $T_{Rule\_in_{S}}$-iteration.

Finally, it is necessary to use consistent $In\_AS_{\mathcal{S}}$ in order to define a sound semantic characterization of the given logic program $P$. This characterization is called the set of atoms {\em derived} in $P$ according to semantics $\mathcal{S}$ according to the following definition.

\begin{df}
\label{derived}
If $In\_AS_{\mathcal{S}}$ is consistent with $Rule\_in_{\mathcal{S}}$, then it is said that $In\_AS_{\mathcal{S}}$ is the set of atoms {\em derived} in $P$ according to semantics $\mathcal{S}$. 
$\Box$
\end{df}

\begin{ex} \label{ex_retazec}
Let a program $P = \{r_1: a\leftarrow , r_2: b\leftarrow\snot a, r_3: c\leftarrow \snot b, r_4: d\leftarrow \snot c\}$ be given.

We get $AF_P = (\ r_1, r_2, r_3, r_4 \}, \{ (r_1, r_2), (r_2, r_3), (r_3, r_4) \})$.
Consider only the preferred semantics. The only preferred extension of $AF_P$ is the set of rules $\{r_1,r_3\}$.\footnote{It is also a stable, grounded and complete extension.}
 We get $\{ \{ r_1, r_3 \} \} = \mathcal{E}_S(AF_P)$, where $S$ is the preferred semantics. It means, $\{r_1,r_3\}$ is the only set of rules, enabled by the preferred semantics according to Definition \ref{enabled}. 

$In\_AS=\{a,c\}$ according to Definition \ref{in-as}. The set of atoms $\{a,c\}$ is consistent with the set of rules $\{r_1,r_3\}$ according to the Definition \ref{consistent}. Finally, according to Definition \ref{derived} is $\{a,c\}$ derived in $P$ according to the preferred semantics.

Notice that this set is the stable model of $P$.
$\Box$
\end{ex}

\begin{ex}
\label{bug}
Consider now a less straightforward example.

Let $P$ be $\{ r_1:a \leftarrow \snot b, \; r_2:b\leftarrow c,\snot d.\}$, $r_3:c \leftarrow.\}$, then $attacks=\emptyset$. If $S$ is the preferred semantics, then  
$\{ \{ r_1, r_2, r_3 \} \} = \mathcal{E}_S(AF_P)$, $P = Rule\_in_{\mathcal{S}}$ is enabled by the preferred semantics.

Further, it holds that $In\_AS_{\mathcal{S}} = \{ a, b, c \}$ according to Definition \ref{in-as}. But $In\_AS_{\mathcal{S}}$ is not consistent with $P = Rule\_in_{\mathcal{S}}$, hence no atom is derived in $P$ according to the preferred semantics.

Consistency checks are intended as a guard against hidden attacks, as our example demonstrates. This is why the set $In\_AS_{\mathcal{S}}$ is not derivable in $P$ according to the preferred semantics. Hence, our construction prevent to accept inconsistent sets of atoms as semantic characterizations of logic programs.

On the other hand, $\{ r_2,r_3 \}$ (may be, also $\{ r_1, r_3 \}$) could be an intuitive preferred extension of an argumentation framework assigned to $P$. It means that our construction do not generate all intuitive semantic characterizations of a logic program corresponding to an argumentation semantics.
$\Box$
\end{ex}

\begin{rk}
\label{wayOUt}
May be, a way out of this bug could be built over subsets of $Rule\_in_{\mathcal{S}}$ and/or of $In\_AS$. Definition \ref{derived} can be modified accordingly as follows: Let $M$ be a maximal subset of $In\_AS_{\mathcal{S}}$ and $R$ be a maximal subset of $Rule\_in_{\mathcal{S}}$ s.t. $M$ is  consistent with $R$. Then it is said that $M$ is the set of atoms {\em derived} in $P$ according to semantics $\mathcal{S}$. 

If we consider Example \ref{bug}, we get sets $\{a, c \}$ and $\{ b, c \}$ as derived atoms corresponding to the preferred extension. However, this is not appropriate for stable semantics. A nice uniform transfer of an argumentation semantics to a logic program semantics would be lost, if a special handling of inconsistency for different argumentation semantics' is specified.

More comments about some possible ways how to fix this bug are included into Section \ref{future}.
$\Box$
\end{rk}

We repeat that the given construction of an argumentation framework over a logic program is useful for goals of Section \ref{repre}. Possibilities of more general constructions aiming at a transfer of an argumentation semantics to a logic program semantics are  presented in Section \ref{future}.

Derivation of atoms according to Definition \ref{derived} coincides with the derivation of derivation in Section \ref{prelim}.

\begin{pr}
Let an argumentation semantics $S$ be given. Let be $R = Rule\_in_S$.
A set of atoms derived in $P$ according to the semantics $S$ is $\Delta^{\leadsto_R}$ for some $\Delta$.
\end{pr}
\noindent
Proof:\\
Let be $R = Rule\_in_{S}$ and $In\_AS$ be the corresponding derived set of atoms.

Suppose that $\Delta = \{ \snot A \mid \exists r \in R \; A \in \sbody^-(r) \}$. It holds that $A \in \Delta^{\leadsto^{R}}$ iff $R^{+}_\Delta \models A$. Obviously, $R^{+}_\Delta \models A$ holds iff $A \in In\_AS$.
$\Box$

An open problem is, how semantics' transferred from argumentation frameworks are related to known semantics of logic programs (stable model semantics, partial stable model semantics, well founded semantics etc.)

Note that stable extensions of $AF_P$ are not in general stable models of $P$.
\begin{ex}
Consider the program $P=\{r_1:a\leftarrow p, \snot b, r_2:b\leftarrow q, \snot a, r_3:p\leftarrow\}$. 

The stable model of $P$ is $\{p,a\}$, but the stable extension of $AF_P$ does not exist, rules $r_1,r_2,r_3$ are mutually conflict-free, but $In\_AS = \{ a, b, p \}$ is not consistent with $Rule\_in = \{ r_1, r_2, r_3 \}$,
$\Box$
\end{ex}

This observation is a consequence of the given design decision concerning the attack relation -- attacking rules are only rules with empty positive part of the body.

\section{Odd cycles}

In this section some examples are presented in order to show that a transfer of an argumentation semantics to a logic program (without a suitable ``classic'' semantic characterization) enables a reasonable semantic characterization of the program.

Some logic programs without stable models have a clear intuitive meaning. A transfer of argumentation semantics from the corresponding argumentation framework enables to catch a meaning of such programs. Of course, a more detailed analysis is needed, in order to understand the relations of those semantics to partial stable models semantics and well founded semantics (or other semantics' of logic programs).\footnote{Some results are presented in the literature, see Section \ref{related}.}

\begin{ex}
Remind Example \ref{ex_retazec}. Let $P^{\prime}$ be $P \cup \{r5:e\leftarrow \snot e\}$. $P^{\prime}$ has no stable model.

The graph of the argumentation framework $AF_{P^{\prime}}$ contains an isolated vertex  $r5$ which attacks itself. If we transfer preferred and grounded semantics from $AF_{P^{\prime}}$ to logic program $P^{\prime}$, we obtain a semantic characterization by an intuitive set of rules $\{r_1,r_3\}$ and, consequently, of atoms $\{a,c\}$ as in Example \ref{ex_retazec}.
$\Box$
\end{ex}

However, a special interest deserves the problem of odd cycles. In this case a transfer from argumentation semantics' to logic program semantics' provides a new perspective on logic programs.\footnote{We realize that this is a complex problem and diverse intuitions should be analyzed.}

\begin{ex}
Consider program $P_1=\{r_1:a\leftarrow \snot b, r_2:b\leftarrow \snot a\}$ with an even (negative) cycle and $P_2=\{r_1:a\leftarrow \snot b, r_2:b\leftarrow \snot c, r_3:c\leftarrow \snot a\}$ with an odd (negative) cycle. There is no stable model of $P_2$.

Preferred, stable and complete argumentation semantics' assign two extensions to $AF_{P_1}$. On the other hand, they assign one (empty) or no extension to $AF_{P_2}$.

Recursive semantics' proposed in \cite{baroni} overcome this asymmetry. Note that $AF_P$ consists of the only component, the odd cycle $(r_1,r_2), (r_2, r_3),(r_3, r_1)$. $CF1$ assigns three extensions $\{\{a\}, \{b\}, \{c\}\}$ to this framework. Our construction enables to transfer this semantics to the logic program $P_2$.
$\Box$
\end{ex}

Consider also other example.

\begin{ex}
Let be $P=\{r_1:a\leftarrow \snot a, r_2:b\leftarrow \snot a\}$. The argumentation framework $AF_P$ has according to the semantics $CF2$ extension $r_2$, consequently $\{b\}$ is transferred to $P$.
$\Box$
\end{ex}

\section{Representation of argumentation framework by logic program}
\label{repre}

In this section we apply a changed view. An argumentation framework $AF$ is assumed and its representation by a simple logic program $P_{AF}$ is constructed. Then we can construct an argumentation framework $\mathcal A$ over the rules of that program using the method of Section \ref{transAF}. Suppose that an argumentation semantics $S$ is applied to the argumentation framework $\mathcal A$ over the rules of the program $P_{AF}$. We will show that an application of transferred argumentation semantics to the logic program $P_{AF}$ produces the same result as the application of  the semantics to the original argumentation framework $AF$.

\begin{df}
Let an argumentation framework $AF=\langle AR,attacks \rangle$ be given. We represent $AF$ by a logic program $P_{AF}$ as follows
\begin{itemize}
\item for each $a \in AR$ there is exactly one rule $r \in P_{AF}$ s.t. $\shead(r)=\{a\}$
\item $\sbody^-(r)=\{ b \mid b\in AR, (b,a) \in attacks\}$, $\sbody^+(r) = \emptyset$.
\end{itemize}
$\Box$
\end{df}

A remark: if body of a rule is empty, then the corresponding argument is not attacked in $AF$.

\begin{ex}
Let be $AF =(AR,attacks)$, where $AR = \{a,b,c,d,e\}$ \\ 
and $attacks=\{(a,b),(c,b),(c,d),(d,c),(d,e),(e,e)\}$. $P_{AF}$, the logic program representing $AF$ is as follows:

\begin{eqnarray*}
r_1:& & b \leftarrow \snot a, \snot c \\
r_2: & & a \leftarrow \\
r_3: & & c \leftarrow \snot d \\
r_4: & & d \leftarrow \snot c \\
r_5: & & e \leftarrow \snot e, \snot d
\end{eqnarray*} 
$\Box$
\end{ex}

Programs representing an argumentation framework look like lists: to each argument in the head of a rule is assigned a list of arguments attacking the argument in the head of the rule. 

Notice that there are logic programs, which cannot represent an argumentation framework. On the other hand, if a logic program represents an argumentation framework, it is done in a unique way -- there is exactly one argumentation framework represented by the program.

\begin{ex}
$P_1 = \{ a\leftarrow \snot b, b\leftarrow \snot a\}$ is a logic program, which represents the argumentation framework $AF=\langle \{a,b\}, \{(a,b),(b,a)\}\rangle$.

$P_2 = \{a\leftarrow \snot b\}$ cannot be a representation of any argumentation framework. There is no rule in $P_2$ with $b$ in its head (and each argument must be in the head of a rule).   
\end{ex} 

\begin{te}
Let $AF$ be an argumentation framework, $AF = (AR, attacks)$, $P_{AF}$ be the logic program representing $AF$. Let $In\_AS$ be a set of atoms, derivable in $P_{AF}$ according to a semantics $S$. 

Then $In\_AS$ is an extension of $AF$ according to the semantics $S$. 
\end{te}

\noindent
Proof:\\
For each argument $a \in AR$, there is exactly one rule $r \in P_{AF}$ s.t. $\shead(r) = a$. A function $\Psi:R\rightarrow AR$, where $R \subseteq P$, assigns to each rule    $r\in R$ the­ argument $a\in AR$, which occurs in the head of $r$. $\Psi^{-1}:AR \rightarrow R$ is an inverse function which assigns to an argument the rule with the argument in the head.

$In\_AS = \{a\mid \exists r \in Rule\_in, head(r)=a\}$ follows from the fact that $\sbody^+(r) = \emptyset$ for each rule $r$. Hence, $In\_AS = \Psi(Rule\_in)$. 

It follows from the definition that for each $(a,b)\in attacks$ there is a pair $(r_1,r_2)\in attacks_P$, where $AF_P = \langle AR_P,attacks_P\rangle$. Notice that $a \in \shead(r_1)$ and in $\shead(r_2)$ is $b$. ($AF_P$ is a framework over the rules of the  program $P$). If $(a,b) \in attacks$ then $\snot a$ occurs in the body of a rule with $b$ in the head. Similarly,  for all $(r_1,r_2)\in attacks_P$ there is $(x,y)\in AF$ s.t.  $\shead(r_1)=x, y\in body^-(r_2)$. Therefore, the only difference between the frameworks  $AF$ and $AF_P$ is that the vertices of both frameworks are renamed according to the function $\Psi$. 

Therefore, $In\_AS=\Psi(Rule\_in)=\mathcal{E}_S(AF)$.

\section{Future goals}
\label{future}

In this section three possible alternative  transfers of argumentation semantics' to logic program semantics' are sketched. Only very preliminary remarks are presented.

\paragraph{Canonical program.}

The first possibility, which we will investigate is as follows.
Suppose, that an argumentation framework is given. We can represent the argumentation framework by a logic program $P_{AF}$ defined in Section \ref{repre} or by its more limpid, straightforward copy $P^{AF}$ defined below.

\begin{df}
Let $AF = (AR, attacks)$ be an argumentation framework.
The logic program $P^{AF}$ assigned to $AF$ is the least set of rules satisfying the conditions:
\begin{itemize}
\item $AR$ is the set of atoms of $P^{AF}$,
\item if $(a, b) \in attacks$, then $(a \leftarrow \snot b) \in P^{AF}$,
\item if $a \in AR$ and neither $(a, b) \in attacks$, nor $(b, a) \in atacks$ for some $b$,
then $(a \leftarrow ) \in P^{AF}$.
\end{itemize}
$\Box$
\end{df}

It can be said, that $P^{AF}$ is the {\em canonical} logic program w.r.t. $AF$. An argumentation semantics of $AF$ can be transferred to a semantics of the canonical program in a rather straightforward way (in terms of dependencies on hypotheses). The planned next step is a transfer of those dependencies to arbitrary logic programs (for some argumentation semantics' a similar work is done by \cite{bondarenko}).

\paragraph{Hypotheses as arguments.}

Dung in his seminal paper \cite{dung} proposed a representation of a logic program in an argumentation framework.  Pairs of the form $(\Delta, A)$, where $\Delta$ is a hypothesis and $A \in \Delta^{\leadsto_P}$ are arguments in \cite{dung}.\footnote{But in \cite{dung2} arguments are hypotheses, too.}

While Dung was focused on expressing a logic program as an argumentation framework, our goal is to transfer argumentation semantics ``back'' to the logic program. An interesting contribution could be a transfer of AD1, AD2, CF1, CF2 and other new semantics specified for $AF^{P}$ back to $P$. 
We will use some notions of \cite{dt} in order to present a similar idea how to consider hypotheses as arguments.

\begin{df}[\cite{dt}]
\label{dtatt}
A hypothesis $\Delta$ attacks another hypothesis $\Delta^{\prime}$ in a program $P$ if there is $A \in \Delta^{\leadsto_P}$ s.t. $\snot A \in \Delta^{\prime}$.

A hypothesis $\Delta$ is self-consistent in $P$, if it does not attack itself
$\Box$
\end{df}

\begin{df}
Let a program $P$ be given. Let $\mathcal H$ be the set of all hypothesis over the language of $P$.

Then an associated argumentation framework $AF^{P} = (AR, attacks)$ is defined as follows. $AR$ is the set of all self-consistent hypotheses of $\mathcal H$ and $attacks$ is defined as in Definition \ref{dtatt}.

If $E \in {\mathcal E}_S(AF^{P})$ for a semantics $S$, then for each $\Delta \in E$ the set of atoms $\Delta^{\leadsto_P}$ provides a semantic characterization of $P$ according to $S$
$\Box$
\end{df}

Notice that this construction is computationally more demanding -- $AF^{P}$ cannot be constructed by an inspection of the syntactic form of $P$. 

Moreover, it is possible that to an extension $E$ of ${\mathcal E}_S(AF^{P})$ is assigned a set of sets of atoms of $P$. It seems that only maximal (w.r.t set-theoretic inclusion) hypotheses of $E$ should be considered if e.g. preferred semantics is transferred.

If we consider Example \ref{bug}, which illustrates a counterintuitive properties of $AF_P$, constructed in Section \ref{transAF}, we get an intuitive solution.

\begin{ex}
Let $P$ be as in Example \ref{bug}. Then $AR$ of $AF^{P}$, the set of self-consistent hypotheses in $P$ is $\{ \emptyset, \{ \snot a \}, \{ \snot b \}, \{ \snot d \}, \{ \snot a, \snot d \} \}$
\\ and $attacks = \{ (\{ \snot d \}, \{ \snot b \}), (\snot a, \snot d \}, \{ \snot b \})$. 

We get that $E = \{ \emptyset, \{ \snot a \}, \{ \snot d \}, \{ \snot a, \snot d \} \}$ is a preferred extension. If only maximal hypotheses are considered, the set of atoms $\{ b, c \}$ is the transferred semantic characterization of $P$. Otherwise, both $\{ c \}$ and $\{ b, c \}$ correspond to $E$.
$\Box$
\end{ex}

We have to study the details and consequences of the presented proposal. 

\paragraph{Derivation of arguments}

A bug caused by assumption $\sbody^+(r) = \emptyset$ in Definition \ref{zaklDef} can be fixed using the approach of \cite{warPref}. Basic argumentation structures and basic attacks are assumed. Basic argumentation structures contain also conditional arguments. A kind of unfolding of conditional arguments is possible thanks to derivation rules, which enable to derive (non-basic) argumentation structures. Similarly, other derivation rules enable derivation of attacks between general argumentation structures. This machinery enables to leave out the condition $\sbody^+(r) = \emptyset$ of Definition \ref{zaklDef}.

\section{Related work}		\label{related}

This section contains only some sketchy remarks, a more detailed analysis and comparison is planned. 

We are familiar with the following types of results: a correspondence of an argumentation semantics and a logic program semantics is described, particularly, a characterization of extensions of abstract argumentation framework in terms of answer sets or other semantics' of logic programs. Encoding extensions of argumentation frameworks in answer set programming is another type of research. Some researchers construct a new semantics of logic programs, inspired by extensions of argumentation frameworks. This goal is close to ours. However, every result about relations between an argumentation semantics and logic program semantics is helpful for our future research.

Some remarks concerning Dung's approach were presented in previous section.

Relations between the ``classic'' argumentation semantics' and corresponding semantic views on logic programs is studied in  \cite{bondarenko}. Of course, the problem of odd cycles is not tackled in the paper. Our future goal is a detailed comparison of constructions of \cite{bondarenko} and ours.

Argumentation framework is constructed and studied in terms of logic programs in \cite{prak}. Arguments are expressed in a logic programming language,
conflicts between arguments are decided with the help of priorities on rules.

A theory of argumentation that can deal with contradiction within an argumentation
framework was presented in \cite{vermeir}. The results was applied to logic programming semantics. A new semantics of logic programs was proposed. The goal is similar as ours, we will devote an attention to this result.

The correspondence between complete extensions in abstract argumentation and
3-valued stable models in logic programming was studied in \cite{caminada}. 

The project "New Methods for Analyzing, Comparing, and Solving Argumentation Problems", see, e.g., \cite{eggly,woltran,cf2woltran}, is focused on implementations of argumentation frameworks in Answer-Set Programming, but also other fundamental theoretical questions are solved. CF2 semantics is studied, too. An Answer Set Programming Argumentation Reasoning Tool (ASPARTIX) is evolved.  

The Mexican group \cite{osorio1,osorio2,osorio3,osorio4,osorio5,osorio6} contributes to research on relations of logic programing and argumentation frameworks, too.  Their attention is devoted to characterizations of argumentation semantics' in terms of logic programming semantics'. Also a characterization of CF2 is provided in terms of answer set models or stratified argumentation semantics, which is based on stratified minimal models of logic programs.

Our main goal, in the context of presented remarks, is to ``import'' semantics' from argumentation frameworks to logic programs. However, results about relations of both areas are relevant for us.

\section{Conclusions}

A method for transferring an arbitrary argumentation semantics to a logic program semantics was developed. The method consists in defining an argumentation framework over the rules of a program. Extensions of the argumentation framework are sets of rules. A set of consequences of those rules is an interpretation, which provides the corresponding semantic characterization of the program.

This method allows a semantic characterization of programs with odd-length (negative) cycles. 
If a simple program is assigned to an argumentation framework, extensions of the original framework and the framework over the rules of that program coincide.

The presented method prevents generation of inconsistent sets of atoms. On the other hand, it does not create sometimes a semantic characterization of the original program, even if there is an intuitive possibility to specify the semantics. Some ways of solving this bug are sketched in the paper.

Open problems, future goals and connections to related work are discussed in previous sections.

{\bf Acknowledgements:} We are grateful to anonymous referees for valuable comments and proposals. This paper was supported by the grant 1/0689/10 of VEGA.



\end{document}